\input epsf
\input amssym
\catcode`\@11\relax
\newif\ify@autoscale \y@autoscaletrue \def\Yautoscale#1{\ifnum #1=0
  \y@autoscalefalse\else\y@autoscaletrue\fi}
\newdimen\y@b@xdim
\newdimen\y@boxdim \y@boxdim=13pt
\def\Yboxdim#1{\y@autoscalefalse\y@boxdim=#1}
\newdimen\y@linethick    \y@linethick=.3pt
\def\Ylinethick#1{\y@linethick=#1}
\newskip\y@interspace \y@interspace=0ex plus 0.3ex
\def\Yinterspace#1{\y@interspace=#1}
\newif\ify@vcenter   \y@vcenterfalse
\def\Yvcentermath#1{\ifnum #1=0 \y@vcenterfalse\else\y@vcentertrue\fi}
\newif\ify@stdtext   \y@stdtextfalse
\def\Ystdtext#1{\ifnum #1=0 \y@stdtextfalse\else\y@stdtexttrue\fi}
\newif\ify@enable@skew   \y@enable@skewfalse
\expandafter\ifx\csname enableskew\endcsname\relax
 \y@enable@skewfalse \else \y@enable@skewtrue\fi
\def\y@vr{\vrule height0.8\y@b@xdim width\y@linethick depth 0.2\y@b@xdim}
\def\y@emptybox{\y@vr\hbox to \y@b@xdim{\hfil}}
\ify@enable@skew
 \def\y@abcbox#1{\if :#1\else
   \y@vr\hbox to \y@b@xdim{\hfil#1\hfil}\fi}
 \def\y@mathabcbox#1{\if :#1\else
   \y@vr\hbox to \y@b@xdim{\hfil$#1$\hfil}\fi}
\else
 \def\y@abcbox#1{\y@vr\hbox to \y@b@xdim{\hfil#1\hfil}}
 \def\y@mathabcbox#1{\y@vr\hbox to \y@b@xdim{\hfil$#1$\hfil}}
\fi
\def\y@setdim{%
  \ify@autoscale%
   \ifvoid1\else\typeout{Package youngtab: box1 not free! Expect an
     error!}\fi%
   \setbox1=\hbox{A}\y@b@xdim=1.6\ht1 \setbox1=\hbox{}\box1%
  \else\y@b@xdim=\y@boxdim \advance\y@b@xdim by -2\y@linethick
  \fi}
\newcount\y@counter
\newif\ify@islastarg
\def\y@lastargtest#1,#2 {\if\space #2 \y@islastargtrue
  \else\y@islastargfalse\fi}
\def\y@emptyboxes#1{\y@counter=#1\loop\ifnum\y@counter>0
  \advance\y@counter by -1 \y@emptybox\repeat}
\def\y@nelineemptyboxes#1{%
  \vbox{%
    \hrule height\y@linethick%
    \hbox{\y@emptyboxes{#1}\y@vr}
    \hrule height\y@linethick}\vskip-\y@linethick}
\def\yng(#1){%
  \y@setdim%
  \hskip\y@interspace%
  \ifmmode\ify@vcenter\vcenter\fi\fi{%
  \y@lastargtest#1,
  \vbox{\offinterlineskip
    \ify@islastarg
     \y@nelineemptyboxes{#1}
    \else
     \y@ungempty(#1)
    \fi}}\hskip\y@interspace}
\def\y@ungempty(#1,#2){%
  \y@nelineemptyboxes{#1}
  \y@lastargtest#2,
  \ify@islastarg
   \y@nelineemptyboxes{#2}
  \else
   \y@ungempty(#2)
  \fi}
\def\y@nelettertest#1#2. {\if\space #2 \y@islastargtrue
  \else\y@islastargfalse\fi}
\def\y@abcboxes#1#2.{%
  \ify@stdtext\y@abcbox#1\else\y@mathabcbox#1\fi%
  \y@nelettertest #2.
  \ify@islastarg\unskip%
   \ify@stdtext\y@abcbox{#2}\else\y@mathabcbox{#2}\fi%
  \else\y@abcboxes#2.\fi}
 \newdimen\y@full@b@xdim
 \newcount\y@m@veright@cnt
\ify@enable@skew
 \def\y@get@m@veright@cnt#1#2.{%
   \if :#1 \advance\y@m@veright@cnt by 1\y@get@m@veright@cnt#2.\fi}
 \let\y@setdim@=\y@setdim
 \def\y@setdim{%
   \y@setdim@ \y@full@b@xdim=\y@b@xdim
   \advance\y@full@b@xdim by 1\y@linethick}
 \def\y@m@veright@ifskew#1{
   \y@m@veright@cnt=0 \y@get@m@veright@cnt#1.
   \moveright \y@m@veright@cnt\y@full@b@xdim}
\else
 \def\y@m@veright@ifskew#1{}
\fi
\def\y@nelineabcboxes#1{%
  \y@nelettertest #1.
  \ify@islastarg
   \y@m@veright@ifskew{#1}
    \vbox{
      \hrule height\y@linethick%
      \hbox{\ify@stdtext\y@abcbox#1\else\y@mathabcbox#1\fi\y@vr}
      \hrule height\y@linethick}\vskip-\y@linethick
  \else
   \y@m@veright@ifskew{#1}
    \vbox{
      \hrule height\y@linethick%
      \hbox{\y@abcboxes #1.\y@vr}%
      \hrule height\y@linethick}\vskip-\y@linethick
  \fi}
\def\young(#1){%
  \y@setdim%
  \hskip\y@interspace%
  \y@lastargtest#1,
  \ifmmode\ify@vcenter\vcenter\fi\fi{%
  \vbox{\offinterlineskip
    \ify@islastarg\y@nelineabcboxes{#1}%
    \else\y@ungabc(#1)%
    \fi}}\hskip\y@interspace}
\def\y@ungabc(#1,#2){%
  \y@nelineabcboxes{#1}%
  \y@lastargtest#2,
  \ify@islastarg\y@nelineabcboxes{#2}%
  \else\y@ungabc(#2)%
  \fi}
\catcode`\@12\relax
 


\def \IntroSection {{\oldsize 1}}
\def \FourZeroSection {{\oldsize 2}}
\def \DynkinFigure {{\oldsize 1}}
\def \CohomologyTable {{\oldsize 1}}
\def \ThreeOneSection {{\oldsize 3}}
\def \GravityTable {{\oldsize 2}}
\def \GravitinoTable {{\oldsize 3}}
\def \Conclusions {{\oldsize 4}}

\Yvcentermath1
\Yboxdim8pt

\newfam\scrfam
\batchmode\font\tenscr=rsfs10 \errorstopmode
\ifx\tenscr\nullfont
        \message{rsfs script font not available. Replacing with calligraphic.}
        \def\scr{\cal}
\else   
        \font\sevenscr=rsfs7
        \font\fivescr=rsfs5
        \skewchar\tenscr='177 \skewchar\sevenscr='177 \skewchar\fivescr='177
        \textfont\scrfam=\tenscr \scriptfont\scrfam=\sevenscr
        \scriptscriptfont\scrfam=\fivescr
        \def\scr{\fam\scrfam}
        \def\cal{\scr}
\fi
\catcode`\@=11
\newfam\frakfam
\batchmode\font\tenfrak=eufm10 \errorstopmode
\ifx\tenfrak\nullfont
        \message{eufm font not available. Replacing with italic.}
        \def\frak{\it}
\else
	
	\font\sevenfrak=eufm7 \font\fivefrak=eufm5
        
	\textfont\frakfam=\tenfrak
	\scriptfont\frakfam=\sevenfrak \scriptscriptfont\frakfam=\fivefrak
	\def\frak{\fam\frakfam}
\fi
\catcode`\@=\active
\newfam\msbfam
\batchmode\font\twelvemsb=msbm10 scaled\magstep1 \errorstopmode
\ifx\twelvemsb\nullfont\def\Bbb{\bf}

	\message{Blackboard bold not available. Replacing with boldface.}
\else   \catcode`\@=11
        \font\tenmsb=msbm10 \font\sevenmsb=msbm7 \font\fivemsb=msbm5
        \textfont\msbfam=\tenmsb
        \scriptfont\msbfam=\sevenmsb \scriptscriptfont\msbfam=\fivemsb
        \def\Bbb{\relax\expandafter\Bbb@}
        \def\Bbb@#1{{\Bbb@@{#1}}}
        \def\Bbb@@#1{\fam\msbfam\relax#1}
        \catcode`\@=\active

\fi
        \font\fivemi=cmmi5
        \font\sixmi=cmmi6
        \font\eightrm=cmr8              \def\xrm{\eightrm}
        \font\eightbf=cmbx8             \def\xbf{\eightbf}
        \font\eightit=cmti10 at 8pt     \def\xit{\eightit}

        \font\eighttt=cmtt8

        \font\eightcp=cmcsc8    
                      \def\xold{\eighti}
        \font\eightmi=cmmi8
                     \def\xbold{\eightib}
        \font\teni=cmmi10               \def\old{\teni}
        \font\tencp=cmcsc10

        \font\twelvecp=cmcsc10 scaled\magstep1
        
        \font\sixrm=cmr6
        \font\fiverm=cmr5

        \font\eightsy=cmsy8
        \font\sixsy=cmsy6
        \font\eightsl=cmsl8

        \font\sixbf=cmbx6

	 at10pt	
	\font\twelvehelvbold=phvb at12pt
	 at14pt
	\font\sixteenhelvbold=phvb at16pt
	 at16pt



\def\xbold{\xbf}
\def\xold{\xrm}


\def\noblackbox{\overfullrule=0pt}
\noblackbox

\def\eightpoint{
\def\rm{\fam0\eightrm}
\textfont0=\eightrm \scriptfont0=\sixrm \scriptscriptfont0=\fiverm
\textfont1=\eightmi  \scriptfont1=\sixmi  \scriptscriptfont1=\fivemi
\textfont2=\eightsy \scriptfont2=\sixsy \scriptscriptfont2=\fivesy
\textfont3=\tenex   \scriptfont3=\tenex \scriptscriptfont3=\tenex
\textfont\itfam=\eightit \def\it{\fam\itfam\eightit}
\textfont\slfam=\eightsl \def\sl{\fam\slfam\eightsl}
\textfont\ttfam=\eighttt \def\tt{\fam\ttfam\eighttt}
\textfont\bffam=\eightbf \scriptfont\bffam=\sixbf 
                         \scriptscriptfont\bffam=\fivebf
                         \def\bf{\fam\bffam\eightbf}
\normalbaselineskip=10pt}



\newtoks\headtext
\headline={\ifnum\pageno=1\hfill\else
	\ifodd\pageno
        \noindent{\eightcp\the\headtext}{ }\dotfill{ }{\old\folio}
	\else\noindent{\old\folio}{ }\dotfill{ }{\eightcp\the\headtext}\fi
	\fi}
\def\makeheadline{\vbox to 0pt{\vss\noindent\the\headline\break
\hbox to\hsize{\hfill}}
        \vskip2\baselineskip}
\newcount\infootnote
\infootnote=0
\newcount\footnotecount
\footnotecount=1
\def\foot#1{\infootnote=1
\footnote{${}^{\the\footnotecount}$}{\vtop{\baselineskip=.75\baselineskip
\advance\hsize by
-\parindent{\eightpoint\rm\hskip-\parindent
#1}\hfill\vskip\parskip}}\infootnote=0\global\advance\footnotecount by
1\hskip2pt}
\newcount\refcount
\refcount=1
\newwrite\refwrite
\def\oldsize{\ifnum\infootnote=1\xold\else\old\fi}
\def\ref#1#2{
	\def#1{{{\oldsize\the\refcount}}\ifnum\the\refcount=1\immediate\openout\refwrite=refs.tex\fi\immediate\write\refwrite{\item{[{\xold\the\refcount}]} 
	#2\hfill\par\vskip-2pt}\xdef#1{{\noexpand\oldsize\the\refcount}}\global\advance\refcount by 1}
	}
\def\refout{\eightpoint\catcode`\@=11
        \xrm\immediate\closeout\refwrite
        \vskip2\baselineskip
        {\noindent\twelvecp References}\hfill\vskip\baselineskip
        \baselineskip=.75\baselineskip
        \input refs.tex
        \baselineskip=4\baselineskip \divide\baselineskip by 3
        \catcode`\@=\active\rm}

\def\skipref#1{\hbox to15pt{\phantom{#1}\hfill}\hskip-15pt}

\def\hepth#1{\href{http://xxx.lanl.gov/abs/hep-th/#1}{arXiv:\allowbreak
hep-th\slash{\xold#1}}}

\def\arxiv#1#2{\href{http://arxiv.org/abs/#1.#2}{arXiv:\allowbreak
{\xold#1}.{\xold#2}}} 
 
\def\jhep#1#2#3#4{\href{http://jhep.sissa.it/stdsearch?paper=#2\%28#3\%29#4}{J. High Energy Phys. {\xbold #1#2} ({\xold#3}) {\xold#4}}}

\def\FP#1#2#3{Fortsch. Phys. {\xbold#1} ({\xold#2}) {\xold#3}}

\def\IJMPA#1#2#3{Int. J. Mod. Phys. {\xbf A}{\xbold#1} ({\xold#2}) {\xold#3}}

\def\MPLA#1#2#3{Mod. Phys. Lett. {\xbf A}{\xbold#1} ({\xold#2}) {\xold#3}}

\def\NPB#1#2#3{Nucl. Phys. {\xbf B}{\xbold#1} ({\xold#2}) {\xold#3}}

\def\PRD#1#2#3{Phys. Rev. {\xbf D}{\xbold#1} ({\xold#2}) {\xold#3}}

\newcount\sectioncount
\sectioncount=0
\def\section#1#2{\global\eqcount=0
	\global\subsectioncount=0
        \global\advance\sectioncount by 1
	\ifnum\sectioncount>1
	        \vskip2\baselineskip
	\fi
\noindent{\twelvecp\the\sectioncount. #2}\par\nobreak
       \vskip.5\baselineskip\noindent
        \xdef#1{{\old\the\sectioncount}}}
\newcount\subsectioncount
\def\subsection#1#2{\global\advance\subsectioncount by 1
\vskip.75\baselineskip\noindent\line{\tencp\the\sectioncount.\the\subsectioncount. #2\hfill}\nobreak\vskip.4\baselineskip\nobreak\noindent\xdef#1{{\old\the\sectioncount}.{\old\the\subsectioncount}}}
\def\immediatesubsection#1#2{\global\advance\subsectioncount by 1
\vskip-\baselineskip\noindent
\line{\tencp\the\sectioncount.\the\subsectioncount. #2\hfill}
	\vskip.5\baselineskip\noindent
	\xdef#1{{\old\the\sectioncount}.{\old\the\subsectioncount}}}
\newcount\subsubsectioncount
\def\subsubsection#1#2{\global\advance\subsubsectioncount by 1
\vskip.75\baselineskip\noindent\line{\tencp\the\sectioncount.\the\subsectioncount.\the\subsubsectioncount. #2\hfill}\nobreak\vskip.4\baselineskip\nobreak\noindent\xdef#1{{\old\the\sectioncount}.{\old\the\subsectioncount}.{\old\the\subsubsectioncount}}}
\newcount\appendixcount
\appendixcount=0
\def\appendix#1{\global\eqcount=0
        \global\advance\appendixcount by 1
        \vskip2\baselineskip\noindent
        \ifnum\the\appendixcount=1
        {\twelvecp Appendix A: #1}\par\nobreak
                        \vskip.5\baselineskip\noindent\fi
        \ifnum\the\appendixcount=2
        {\twelvecp Appendix B: #1}\par\nobreak
                        \vskip.5\baselineskip\noindent\fi
        \ifnum\the\appendixcount=3
        {\twelvecp Appendix C: #1}\par\nobreak
                        \vskip.5\baselineskip\noindent\fi}
\def\acknowledgements{\immediate\write\contentswrite{\item{}\hbox
        to\contentlength{Acknowledgements\dotfill\the\pageno}}
        \vskip2\baselineskip\noindent
        \underbar{\it Acknowledgements:}\ }
\newcount\eqcount
\eqcount=0
\def\Eqn#1{\global\advance\eqcount by 1
\ifnum\the\sectioncount=0
	\xdef#1{{\noexpand\oldsize\the\eqcount}}
	\eqno({\oldstyle\the\eqcount})
\else
        \ifnum\the\appendixcount=0
\xdef#1{{\noexpand\oldsize\the\sectioncount}.{\noexpand\oldsize\the\eqcount}}
                \eqno({\oldstyle\the\sectioncount}.{\oldstyle\the\eqcount})\fi
        \ifnum\the\appendixcount=1
	        \xdef#1{{\noexpand\oldstyle A}.{\noexpand\oldstyle\the\eqcount}}
                \eqno({\oldstyle A}.{\oldstyle\the\eqcount})\fi
        \ifnum\the\appendixcount=2
	        \xdef#1{{\noexpand\oldstyle B}.{\noexpand\oldstyle\the\eqcount}}
                \eqno({\oldstyle B}.{\oldstyle\the\eqcount})\fi
        \ifnum\the\appendixcount=3
	        \xdef#1{{\noexpand\oldstyle C}.{\noexpand\oldstyle\the\eqcount}}
                \eqno({\oldstyle C}.{\oldstyle\the\eqcount})\fi
\fi}
\def\eqn{\global\advance\eqcount by 1
\ifnum\the\sectioncount=0
	\eqno({\oldstyle\the\eqcount})
\else
        \ifnum\the\appendixcount=0
                \eqno({\oldstyle\the\sectioncount}.{\oldstyle\the\eqcount})\fi
        \ifnum\the\appendixcount=1
                \eqno({\oldstyle A}.{\oldstyle\the\eqcount})\fi
        \ifnum\the\appendixcount=2
                \eqno({\oldstyle B}.{\oldstyle\the\eqcount})\fi
        \ifnum\the\appendixcount=3
                \eqno({\oldstyle C}.{\oldstyle\the\eqcount})\fi
\fi}
\def\multi{\global\advance\eqcount by 1}
\def\multieqn#1{({\oldstyle\the\sectioncount}.{\oldstyle\the\eqcount}\hbox{#1})}
\def\multiEqn#1#2{\xdef#1{{\oldstyle\the\sectioncount}.{\old\the\eqcount}#2}
        ({\oldstyle\the\sectioncount}.{\oldstyle\the\eqcount}\hbox{#2})}
\def\multiEqnAll#1{\xdef#1{{\oldstyle\the\sectioncount}.{\old\the\eqcount}}}
\newcount\tablecount
\tablecount=0
\def\Table#1#2#3{\global\advance\tablecount by 1
\immediate\write\intrefwrite{\def\noexpand#1{{\noexpand\oldsize\the\tablecount}}}
       \vtop{\vskip2\parskip
       \centerline{#2}
       \vskip5\parskip
       {\narrower\noindent\it Table \the\tablecount: #3\smallskip}
       \vskip2\parskip}}
\newcount\figurecount
\figurecount=0
\def\Figure#1#2#3{\global\advance\figurecount by 1
\immediate\write\intrefwrite{\def\noexpand#1{{\noexpand\oldsize\the\figurecount}}}
       \vtop{\vskip2\parskip
       \centerline{#2}
       \vskip4\parskip
       \centerline{\it Figure \the\figurecount: #3}
       \vskip3\parskip}}
\def\TextFigure#1#2#3{\global\advance\figurecount by 1
\immediate\write\intrefwrite{\def\noexpand#1{{\noexpand\oldsize\the\figurecount}}}
       \vtop{\vskip2\parskip
       \centerline{#2}
       \vskip4\parskip
       {\narrower\noindent\it Figure \the\figurecount: #3\smallskip}
       \vskip3\parskip}}
\newtoks\url
\def\Href#1#2{\catcode`\#=12\url={#1}\catcode`\#=\active#2}
\def\href#1#2{{#2}}

\parskip=3.5pt plus .3pt minus .3pt
\baselineskip=14pt plus .1pt minus .05pt
\lineskip=.5pt plus .05pt minus .05pt
\lineskiplimit=.5pt
\abovedisplayskip=18pt plus 4pt minus 2pt
\belowdisplayskip=\abovedisplayskip
\hsize=14cm
\vsize=19cm
\hoffset=1.5cm
\voffset=1.8cm
\frenchspacing
\footline={}
\raggedbottom

\newskip\origparindent
\origparindent=\parindent
\def\ts {\textstyle}
\def\ss{\scriptstyle}

\def\*{\partial}

\def\={\!=\!}
\def\small#1{{\hbox{$#1$}}}

\def\fraction#1{\small{1\over#1}}
\def\fr{\fraction}
\def\Fraction#1#2{\small{#1\over#2}}
\def\Fr{\Fraction}

\def\eg{{\it e.g.}}

\def\ie{{\it i.e.}}

\def\HH{{\Bbb H}}


\def\appendix#1#2{\global\eqcount=0
        \global\advance\appendixcount by 1
        \vskip2\baselineskip\noindent
        \ifnum\the\appendixcount=1
        \immediate\write\intrefwrite{\def\noexpand#1{A}}
        {\twelvecp Appendix A: #2}\par\nobreak
                        \vskip.5\baselineskip\noindent\fi
        \ifnum\the\appendixcount=2
        {\twelvecp Appendix B: #2}\par\nobreak
                        \vskip.5\baselineskip\noindent\fi
        \ifnum\the\appendixcount=3
        {\twelvecp Appendix C: #2}\par\nobreak
                        \vskip.5\baselineskip\noindent\fi}

\def\textfrac#1#2{\raise .45ex\hbox{\the\scriptfont0 #1}\nobreak\hskip-1pt/\hskip-1pt\hbox{\the\scriptfont0 #2}}


\def\frac{\Fr}

\def\mathbb{\Bbb}





\def\so{{\frak so}}
\def\sl{{\frak sl}}
\def\sp{{\frak sp}}
\def\usp{{\frak usp}}
\def\osp{{\frak osp}}


\catcode`@=11
\def\openupnormal{\afterassignment\@penupnormal\dimen@=}
\def\@penupnormal{\advance\normallineskip\dimen@
  \advance\normalbaselineskip\dimen@
  \advance\normallineskiplimit\dimen@}
\catcode`@=12

\def\EqMatrix{\let\quad\enspace\openupnormal6pt\matrix}



\def\textfrac#1#2{\raise .45ex\hbox{\the\scriptfont0 #1}\nobreak\hskip-1pt/\hskip-1pt\hbox{\the\scriptfont0 #2}}


\def\frac{\Fr}

\def\mathbb{\Bbb}

\newskip\scrskip
\scrskip=-.6pt plus 0pt minus .1pt


\newwrite\intrefwrite
\immediate\openout\intrefwrite=superfield.intref

\newwrite\contentswrite

\newdimen\sublength
\sublength=\hsize 
\advance\sublength by -\parindent

\newdimen\contentlength
\contentlength=\sublength

\advance\sublength by -\parindent

\def\section#1#2{\global\eqcount=0
	\global\subsectioncount=0
        \global\advance\sectioncount by 1
\ifnum\the\sectioncount=1\immediate\openout\contentswrite=superfield.contents\fi
\immediate\write\contentswrite{\item{\the\sectioncount.}\hbox to\contentlength{#2\dotfill\the\pageno}}
        \ifnum\sectioncount>1
		        \vskip2\baselineskip
	\fi
\immediate\write\intrefwrite{\def\noexpand#1{{\noexpand\oldsize\the\sectioncount}}}\noindent{\twelvecp\the\sectioncount. #2}\par\nobreak
       \vskip.5\baselineskip\noindent}

\def\subsection#1#2{\global\advance\subsectioncount by 1
\immediate\write\contentswrite{\itemitem{\the\sectioncount.\the\subsectioncount.}\hbox
to\sublength{#2\dotfill\the\pageno}}
\immediate\write\intrefwrite{\def\noexpand#1{{\noexpand\oldsize\the\sectioncount}.{\noexpand\oldsize\the\subsectioncount}}}\vskip.75\baselineskip\noindent\line{\tencp\the\sectioncount.\the\subsectioncount. #2\hfill}\nobreak\vskip.4\baselineskip\nobreak\noindent}

\def\immediatesubsection#1#2{\global\advance\subsectioncount by 1
\immediate\write\contentswrite{\itemitem{\the\sectioncount.\the\subsectioncount.}\hbox
to\sublength{#2\dotfill\the\pageno}}
\immediate\write\intrefwrite{\def\noexpand#1{{\noexpand\oldsize\the\sectioncount}.{\noexpand\oldsize\the\subsectioncount}}}
\vskip-\baselineskip\noindent
\line{\tencp\the\sectioncount.\the\subsectioncount. #2\hfill}
	\vskip.5\baselineskip\noindent}

\def\refout{\eightpoint\catcode`\@=11
        \immediate\write\contentswrite{\item{}\hbox to\contentlength{References\dotfill\the\pageno}}
        \xrm\immediate\closeout\refwrite
        \vskip2\baselineskip
        {\noindent\twelvecp References}\hfill\vskip\baselineskip
        \baselineskip=.75\baselineskip
        \input refs.tex
        \baselineskip=4\baselineskip \divide\baselineskip by 3
        \catcode`\@=\active\rm}



\def\threeplus{\yng(1,1,1)_+}

\headtext={Superfields for D=6 exotic supergravity}


\ref\HullExoticGravity{C.M. Hull, {\xit ``Strongly coupled gravity
and duality''}, \NPB{583}{2000}{237} [\hepth{0004195}].}

\ref\HullExoticGravityII{C.M. Hull, {\xit ``Symmetries and
compactifications of (4,0) conformal
gravity''}, \jhep{00}{12}{2000}{007} [\hepth{0011215}].}

\ref\PureSGI{M. Cederwall, {\xit ``Towards a manifestly supersymmetric
    action for D=11 supergravity''}, \jhep{10}{01}{2010}{117}
    [\arxiv{0912}{1814}].}  

\ref\PureSGII{M. Cederwall, 
{\xit ``D=11 supergravity with manifest supersymmetry''},
    \MPLA{25}{2010}{3201} [\arxiv{1001}{0112}].}

\ref\CederwallKarlssonBI{M. Cederwall and A. Karlsson, {\xit ``Pure
spinor superfields and Born--Infeld theory''},
\jhep{11}{11}{2011}{134} [\arxiv{1109}{0809}].}

\ref\ChiodaroliGunaydinRoiban{M. Chiodaroli, M. G\"unayd\i n and
R. Roiban, {\xit ``Superconformal symmetry and maximal supergravity in
various dimensions''}, \jhep{12}{03}{2012}{093} [\arxiv{1108}{3085}],}

\ref\BerkovitsNonMinimal{N. Berkovits,
{\xit ``Pure spinor formalism as an N=2 topological string''},
\jhep{05}{10}{2005}{089} [\hepth{0509120}].}

\ref\PureSpinorOverview{M. Cederwall, {\xit ``Pure spinor superfields
--- an overview''}, Springer Proc. Phys. {\xbf153} ({\xrm2013}) {\xrm61} 
[\arxiv{1307}{1762}].}

\ref\HenneauxLekeuLeonard{M. Henneaux, V. Lekeu and A. Leonard, {\xit
``The action of the (free) (4,0)-theory''}, \jhep{01}{18}{2018}{273}
[\arxiv{1711}{07448}].}

\ref\HenneauxLekeuMatulichProhaska{M. Henneaux, V. Lekeu, J. Matulich
and S. Prohaska, {\xit
``The action of the (free) (3,1) theory in six spacetime
dimensions''},
\jhep{06}{18}{2018}{057}
[\arxiv{1804}{10125}].}

\ref\CederwallTensorAction{M. Cederwall, unpublished.}

\ref\CederwallBLG{M. Cederwall, {\xit ``N=8 superfield formulation of
the Bagger--Lambert--Gustavsson model''}, \jhep{08}{09}{2008}{116}
[\arxiv{0808}{3242}].}

\ref\CederwallABJM{M. Cederwall, {\xit ``Superfield actions for N=8 
and N=6 conformal theories in three dimensions''},
\jhep{08}{10}{2008}{70}
[\arxiv{0809}{0318}].}

\ref\CederwallReformulation{M. Cederwall, {\xit ``An off-shell superspace
reformulation of $D=4$, $N=4$ super-Yang--Mills theory''},
\FP{66}{2018}{1700082} [\arxiv{1707}{00554}].}

\ref\CederwallDSix{M. Cederwall, {\xit ``Pure spinor superspace action
for D=6, N=1 super-Yang--Mills theory''}, \jhep{18}{05}{2018}{115}
[\arxiv{1712.02284}]}

\ref\CederwallPalmkvistSaberiInProgress{M. Cederwall, J, Palmkvist and
I. Saberi, work in progress.}

\ref\CederwallPalmkvistBorcherds{M. Cederwall and J. Palmkvist, {\xit
``Superalgebras, constraints and partition functions''},
\jhep{08}{15}{2015}{36} [\arxiv{1503}{06215}].}

\ref\EagerSaberiWalcher{R. Eager, I. Saberi and J. Walcher, {\xit
``Nilpotence varieties''}, \arxiv{1807}{03766}.}

\ref\HenneauxLekeuLeonard{M. Henneaux, V. Lekeu and A. Leonard, {\xit
``The action of the (free) (4,0)-theory''}, \jhep{01}{18}{2018}{273}
[\arxiv{1711}{07448}].} 

\ref\HenneauxLekeuMatulichProhazka{M. Henneaux, V. Lekeu, J. Matulich
and S. Prohazka, {\xit
``The action of the (free) N=(3,1) theory in six spacetime
dimensions''}, \jhep{18}{06}{2018}{057} [\arxiv{1804}{10125}].}

\ref\MinasianStrickland{R. Minasian and C. Strickland-Constable, {\xit
``On symmetries and dynamics of exotic supermultiplets''}, \arxiv{2007}{08888}.}

\ref\BertrandHoheneggerHohmSamtleben{Y. Bertrand, S. Hohenegger,
O. Hohm and H. Samtleben, {\xit ``Toward exotic 6D supergravities''}, \arxiv{2007}{11644}.}

\ref\SpinorialCohomology{M. Cederwall, B.E.W. Nilsson and D. Tsimpis, 
{\xit ``Spinorial cohomology and maximally supersymmetric theories''},
\jhep{02}{02}{2002}{009} [\hepth{0110069}].}

\ref\BerkovitsI{N. Berkovits, 
{\xit ``Super-Poincar\'e covariant quantization of the superstring''}, 
\jhep{00}{04}{2000}{018} [\hepth{0001035}].}

\ref\BerkovitsIII{N. Berkovits, 
{\xit ``Cohomology in the pure spinor formalism for the
superstring''}, 
\jhep{00}{09}{2000}{046} [\hepth{0006003}].}

\ref\GunaydinDSix{M. G\"unayd\i n, {\xit ``Unified non-metric (1,0)
tensor-Einstein supergravity theories and (4,0) supergravity in six
dimensions''}, \arxiv{2009}{01374}.}

\ref\Strathdee{J.A. Strathdee, {\xit ``Extended Poincar\'e
supersymmetry''}, \IJMPA{2}{1987}{273}.}

\ref\Movshev{M. Movshev and A. Schwarz, {\xit ``On maximally
supersymmetric Yang--Mills theories''}, \NPB{681}{2004}{324}
[\hepth{0311132}].}

\ref\MovshevSchwarzDef{M. Movshev and A. Schwarz, {\xit
``Supersymmetric deformations of maximally supersymmetric gauge
theories''}, \jhep{12}{09}{2012}{136} [\arxiv{0910}{0620}].}

\ref\HohmSambtlebenI{O.~Hohm and H.~Samtleben,
  {\xit ``Exceptional field theory I: $E_{6(6)}$ covariant form of
  M-theory and type IIB''}, 
  \PRD{89}{2014}{066016} [\arxiv{1312}{0614}].}

\ref\CederwallPalmkvistExtendedGeometry{M. Cederwall and J. Palmkvist,
{\xit ``Extended geometries''}, \jhep{02}{18}{2018}{071} [\arxiv{1711}{07694}].}


%
\line{
\epsfysize=18mm
\epsffile{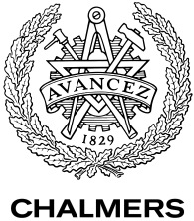}
\hfill}
\vskip-16mm

\line{\hfill}
\line{\hfill Gothenburg preprint}
\line{\hfill December, {\old2020}}
\line{\hrulefill}


\headtext={Cederwall: 
``Superspace formulation of exotic supergravities in six dimensions''}

\vfill

\centerline{\sixteenhelvbold
Superspace formulation}

\vskip3\parskip

\centerline{\sixteenhelvbold
 of exotic supergravities in six dimensions}

%

\vfill

\centerline{\twelvehelvbold Martin Cederwall}

\vfill
\vskip-1cm

\centerline{\it Department of Physics}
\centerline{\it Chalmers University of Technology}
\centerline{\it SE 412 96 Gothenburg, Sweden}

\vfill

{\narrower\noindent \underbar{Abstract:}
We provide a linearised superfield description
of the exotic non-metric $N=(4,0)$
supergravity in $D=6$, by using a pure spinor superfield formalism.
The basic field $\Psi$ is a ghost number 2 scalar, transforming
in the same R-symmetry module as
the tensor fields. Partial results for the $N=(3,1)$ model are presented.
\smallskip}
\vfill

\font\xxtt=cmtt6

\vtop{\baselineskip=.6\baselineskip\xxtt
\line{\hrulefill}
\catcode`\@=11
\line{email: martin.cederwall@chalmers.se\hfill}
\catcode`\@=\active
}

\eject


\catcode`\@=11
        \vskip2\baselineskip
        {\noindent\twelvecp Contents}\hfill\vskip\baselineskip
        \input superfield.contents
        \catcode`\@=\active\rm
\vskip3\baselineskip

\section\IntroSection{Introduction}The existence
of certain exotic gravity-like, but
non-metric,
supermultiplets in six
dimensions was established by Strathdee [\Strathdee] and used by Hull
[\HullExoticGravity,\HullExoticGravityII] to build the physical models.
The models are chiral, with 
$N=(4,0)$ or $(3,1)$ supersymmetry, reduce to ordinary supergravity in
$D=5$, and are suspected to be related to
strong coupling limits of $D=5$ supergravity.
These models have received a renewed interest during the last few years
[\HenneauxLekeuLeonard\skipref\HenneauxLekeuMatulichProhazka\skipref\MinasianStrickland\skipref\GunaydinDSix-\BertrandHoheneggerHohmSamtleben],
concerning for example Lagrangian formulations, supersymmetry and
the possibility to formulate the models in terms of $E_6$ extended geometry
[\HohmSambtlebenI,\CederwallPalmkvistExtendedGeometry].
Interacting versions of these models are still unknown, and should not
be local field theories.

The purpose of the present letter is to present a superfield
formulation. The main focus is on the $N=(4,0)$ model, and partial
results are given for the $N=(3,1)$
model.
The method used is ``pure spinor superfield theory''
[\BerkovitsI\skipref\BerkovitsIII\skipref\SpinorialCohomology\skipref\CederwallBLG\skipref\CederwallABJM\skipref\PureSGI\skipref\PureSGII\skipref\CederwallDSix\skipref\CederwallReformulation\skipref\Movshev\skipref\MovshevSchwarzDef-\PureSpinorOverview]. It
should be stressed that this formalism, which in addition to the
standard superspace coordinates $Z^M=(x^m,\theta^\mu)$ involve
constrained bosonic spinor ghosts, is nothing exotic, or a choice that may
be avoided. Any supermultiplet has a natural formulation in terms of a
pure spinor superfield, and the cohomological equations at a given
ghost number are the standard superspace equations.
However, with the inclusion of a constrained spinor, there is the
advantage of treating all ghost numbers at 
once, \ie, fields, gauge symmetries, equations of motion (for maximal
supersymmetry) etc.
They arise collectively from a single field.
In addition, it often provides a unique way of
writing an action, even for maximally supersymmetric models like
$D=11$ supergravity [\PureSGI,\PureSGII].
In the present case, we are dealing with chiral theories, and the pure
spinor superfield formalism will not yield an action.

\vfill\eject

\section\FourZeroSection{$N=(4,0)$}A superfield formulation
of the $D=6$, $N=(4,0)$ theory
should be based on a ``pure spinor'' $\lambda$ in $(001)(1000)$ of
$D_3\oplus C_4$, the same module as the fermionic coordinates $\theta$.
(We are working in a flat Minkowski background, so the fermionic
variables are spinors.)
It should obey the generic type of constraint
$(\lambda\gamma^a\lambda)=0$, which means that the
remaining modules in $\lambda^2$ are $(002)(2000)\oplus(100)(0100)$.
$\lambda$ is not in a minimal orbit, but an intermediate one; this is
normal, and happens \eg\ in $D=11$ supergravity [\PureSGI,\PureSGII].
We still use the term ``pure spinor'', by need to keep in mind that it
differs from a pure spinor in the sense of Weyl, which by definition
lies in a minimal orbit.
The conventions for the Dynkin labels (\ie, for the ordering of the
nodes in the Dynkin diagram for $D_3\oplus C_4$ is the standard one,
with $(100)(0000)$ being the $Spin(1,5)$ vector. The 32-dimensional
chiral spinors in $(010)(1000)$ and $(001)(1000)$ are real through a
$USp(8)$-Majorana condition, which is most easily seen using a
quaternionic language, where $\so(1,5)\simeq\sl(2;\HH)$ and
$\usp(8)\simeq{\frak a}_4(\HH)$, the algebra of $4\times4$ antihermitean
quaternionic matrices. Then the spinors are $2\times4$ quaternionic
matrices, acted on by $\sl(2;\HH)$ from the left and ${\frak a}_4(\HH)$ from
the right.

\TextFigure\DynkinFigure{\epsffile{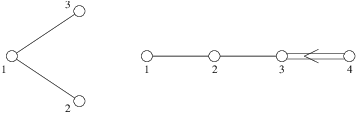}}{The Dynkin diagram, with
our conventions for numbering of the nodes,
of $D_3\oplus C_4\simeq\so(6)\oplus\sp(8)$.}

We want to formulate the on-shell fields as cohomology of the BRST
operator
$$
Q=\lambda^\alpha D_\alpha\eqn
$$
in some field $\Psi(x,\theta,\lambda)$.
Here, $D_\alpha$ is the fermionic covariant derivative, whose torsion
in flat superspace is given by the gamma matrices.  
The BRST operator is nilpotent
due to the constraint on $\lambda$.
However, the pure spinor constraint is irreducible, and removes $6$ of the $32$
degrees of freedom in $\lambda$. This implies that the cohomology of a
scalar field becomes trivial. The zero-mode cohomology (\ie, the
algebraic cohomology when ${\*\over\*x}=0$) only consists of forms,
$$
\Psi_0(\theta,\lambda)=\sum\limits_{n=0}^6
(\lambda\gamma^{a_1}\theta)\ldots(\lambda\gamma^{a_n}\theta)A_{a_1\ldots
a_n}\;.
\Eqn\ZeroModeForms
$$
The full cohomology becomes de Rahm cohomology.
Even if the field $\Psi$ has no local degrees of freedom, it may play
a r\^ole in an interacting theory, like \eg\ is refs.
[\CederwallBLG,\CederwallABJM,\CederwallReformulation].

The partition function of the pure spinor can be given explicitly.
A partition only counting the number of functions at each power of
$\lambda$ is
$$
\eqalign{
Z_\lambda(t)&=1+32\,t+522\, t^2 + 5\,792\, t^3 + 49\,207\, t^4
+ 341\,568\, t^5 + 2\,018\,524\, t^6 + 
 10\,447\,424\, t^7\cr
 &\qquad+ 48\,349\,899\, t^8 + 203\,253\,024\, t^9 + 785\,789\,394\,
 t^{10}+\ldots\cr
 &={(1+t)^6\over(1-t)^{26}}\;,
}\eqn
$$
reflecting the number of degrees of freedom (26) in the pure spinor.
A refined partition function, where the coefficients in the power
expansion take values in the representation ring, is
$$
\eqalign{
{\cal Z}_\lambda(t)
&=
{\bigoplus\limits_{i=0}^6\wedge^i(100)(0000)t^{2i}
\over(1-t)^{(001)(1000)}}\cr
&=(1-t)^{-(001)(1000)}(1-t^2)^{(100)(0000)}\;.
}\eqn
$$
Here, we have used the shorthand ``$(1-t^k)^{\pm R}$'' for the
partition function of a bosonic (minus sign) or fermionic (plus sign)
variable at level $k$.
There are (at least) two ways to read this expression. One is 
to understand the numerator in the first expression as the zero-mode
cohomology, the differential forms of eq. (\ZeroModeForms). The other reads the
last expression as the product of the partition function for an
unconstrained spinor and the partition function of the fermionic ghost
for the constraint in $(100)(0000)$. Since the constraint is
irreducible, there are no further ghosts. 

The latter reading displays the Koszul duality [\Movshev]
of the pure spinor
to a superalgebra ${\cal A}$ through the relation
[\CederwallPalmkvistBorcherds] 
$$
{\cal Z}_\lambda(t)\otimes{\cal Z}_{{\cal A}^+}(t)=1\;.\eqn
$$
Here, ${\cal Z}_{{\cal A}^+}$ is the partition function for the
(universal enveloping algebra of) the positive level subalgebra
${\cal A}^+\subset{\cal A}$.
The superalgebra ${\cal A}$ is a Borcherds superalgebra, but since it
is finite-dimensional (due to finite reducibility), it is also a
finite-dimensional contragredient Lie superalgebra. In the present
case, we obtain
${\cal A}\simeq \osp(8|8)$, the superconformal algebra of an $N=(4,0)$
theory in $D=6$.
Similar appearances of the superconformal algebra seem to be typical
for superconformal systems in a pure spinor superfield approach
[\CederwallBLG,\CederwallABJM,\CederwallReformulation,\CederwallPalmkvistSaberiInProgress].

One needs a field in some non-trivial module. In addition, it should
be subject to some relation involving $\lambda$, so that the
cohomology not just becomes the tensor product of the module of the
field with the (trivial) cohomology of a scalar field.
Here, it is helpful to remember that all fields, including ghosts and
antifields, will appear as zero-mode cohomology. The module of the
field itself will coincide with that of the ghosts with highest ghost
number. In the $(4,0)$ theory, there are $27$ $2$-form tensor fields,
whose ghosts for ghosts are in $(000)(0100)$.
We take a field $\Psi$ in $(000)(0100)$ and also impose
$\Psi\approx\Psi+(\lambda\varrho)|_{(000)(0100)}$, for any
$\varrho(x,\theta,\lambda)$ in $(010)(1000)$. This type of ``shift
symmetry''  is described in detail in ref. [\CederwallKarlssonBI].
In terms of the terminology of ref. [\EagerSaberiWalcher], one would say that
$\Psi$ is not a function on the pure spinor space, but belongs
to a section of a sheaf over pure spinor space.

Now, the calculation  of the zero-mode cohomology is purely algebraic,
and can  be performed  on a computer. The result is displayed in Table
\CohomologyTable.
With a conformal assignment of dimensions, the scalars should have
dimension 2, which means that $\Psi$ has dimension 0.        
The
superfields at different lower numbers are shifted downward, so that fields in
the same row has the same dimension. Since $\lambda$ has dimension
$-\fr2$, the superfield at ghost number $-2+n$ has dimension $\Fr n2$.
Black dots in the table denotes the absence of cohomology.

\Table\CohomologyTable
{$$\hskip-2cm
\vtop{\baselineskip25pt\lineskip0pt
\ialign{
$\hfill#\quad$&$\ss\,\hfill#\hfill\,$&$\ss\,\hfill#\hfill\,$
&$\ss\,\hfill#\hfill\,$&$\ss\,\hfill#\hfill\,$&$\ss\,\hfill#\hfill$
&$\ss\,\hfill#\hfill$&$\ss\,\hfill#\hfill$&\quad#\cr
\hfill\hbox{\eightrm gh\#}=&\ts2&\ts1&\ts0&\ts-1&\ts-2&\ts-3&\ts-4&\cr
&\,\,(000)(0100)\,\, \cr
&(010)(1000)&\bullet&             \cr 
&(020)(0000)&(100)(0100)&\bullet&       &      &\cr
&\bullet&(001)(1000)\,(110)(1000)&\bullet&\bullet&       &\cr
&\bullet&(011)(0000)\,(120)(0000)&(000)(0001)\,(011)(0100)&\bullet&\bullet\cr
&\bullet&\bullet&\raise3pt\vtop{\baselineskip6pt\ialign{
					\hfill$#$\hfill\cr
					\ss(010)(0010)\,(010)(1000)\cr
					\ss(021)(1000)\,(101)(1000)\cr}}
                 &\bullet&\bullet&\bullet\cr
&\bullet&\bullet&\raise3pt\vtop{\baselineskip6pt\ialign{
					\hfill$#$\hfill\cr
					\ss(020)(0000)\,(031)(0000)\cr
					\ss(100)(0000)\,(111)(0000)\cr}}
		&(002)(0100)&\bullet&\bullet&\bullet\cr
&\bullet&\bullet&\bullet&\raise3pt\vtop{\baselineskip6pt\ialign{
					\hfill$#$\hfill\cr
					\ss(001)(0010)\,(001)(1000)\cr
					\ss(012)(1000)\,(110)(1000)\cr}}
                       &\bullet&\bullet&\bullet\cr
&\bullet&\bullet&\bullet&\raise6pt\vtop{\baselineskip6pt\ialign{
					\hfill$#$\hfill\cr
					\ss(000)(0000)\,(000)(0001)\cr
					\ss(011)(0000)\,(022)(0000)\cr
                                        \ss(120)(0000)\,(200)(0000)\cr}}
                       &\bullet&\bullet&\bullet\cr
&\bullet&\bullet&\bullet&\bullet&(010)(1000)\,(101)(1000)&\bullet&\bullet\cr
&\bullet&\bullet&\bullet&\bullet&\raise3pt\vtop{\baselineskip6pt\ialign{
					\hfill$#$\hfill\cr
					\ss(020)(0000)\,(100)(0000)\cr
					\ss(111)(0000)\cr}}&\bullet&\bullet\cr
&\bullet&\bullet&\bullet&\bullet&\bullet&(001)(1000)&\bullet\cr
&\bullet&\bullet&\bullet&\bullet&\bullet&(011)(0000)&\bullet\cr
&\bullet&\bullet&\bullet&\bullet&\bullet&\bullet&\qquad\bullet\qquad\cr
}}
$$}
{The zero-mode cohomology in $\Psi$. The
superfields at different ghost numbers are shifted so that fields in
the same row has the same dimension. Black dots denote the absence of
cohomology.}

\vfill\eject

Coordinate dependence of the fields in these modules will be related
by derivatives 
in the full cohomology. Obviously, fields in different $C_4\approx
USp(8)$ R-symmetry modules do not talk to each other in the linear
theory.
From Table \CohomologyTable, we can thus read the sequence of ghosts
and fields for each of the fields. The action of the BRST operator
(the nilpotent arrows in the following sequences) 
takes us one step to the right and two down.

For the fields in $(0100)={\bf27}$ we read off the sequence
$$
{\bf1}\rightarrow\yng(1)\rightarrow\yng(1,1)\rightarrow
\yng(1,1,1)_-
\eqn
$$
interpreted as the ghost for ghost, the ghost, the $B$-field, and its
antifields or equation of motion $F_-=0$.  Here, the convention is
$$
(020)=\yng(1,1,1)_+\;,\quad(002)=\yng(1,1,1)_-\;.\eqn
$$

The scalars in $(0001)={\bf42}$ have their antifields in the same
module at the correct dimension.

The spinors in $(0010)={\bf48}$ are found at the right dimension, with
their equation of motion.

The ``gravitino'' sequence in $(1000)={\bf8}$ is
$$
\alpha
\rightarrow\yng(1)\otimes\alpha
\rightarrow\yng(1,1)\otimes\alpha
\rightarrow\yng(1,1,1)\otimes\alpha\ominus(030)
\rightarrow\left(\;\yng(1,1,1,1)\otimes\alpha\right)'
\rightarrow\left(\;\yng(1,1,1,1,1)\otimes\alpha\right)'
\eqn
$$
where $\alpha=(010)$, and the ``gravity'' sequence in $(0000)={\bf1}$ is
$$
\threeplus
\rightarrow\yng(1)\otimes\threeplus
\rightarrow\yng(1,1)\otimes\threeplus
\rightarrow\yng(1,1,1)\otimes\threeplus\ominus(040)
\rightarrow\left(\;\yng(1,1,1,1)\otimes\threeplus\right)'
\rightarrow\left(\;\yng(1,1,1,1,1)\otimes\threeplus\right)'
\eqn
$$
In both these cases, the equations of motion are the exterior derivative
on the fields, except for the modules $(030)$ and $(040)$,
respectively, which instead are surviving gauge invariant field strengths (this
is indeed true also for the tensor fields, where the invariant field
strength is in $(020)(0100)$).
The primed tensor product in the ``syzygies'' (the reducibilities of
the equations  of motion, in two steps) implies a projection on
the irreducible modules which are not reached via $(030)$  or $(040)$,
respectively, and the arrow is nilpotent.
So, by definition,
$$
\eqalign{
\left(\yng(1,1,1)\otimes\alpha\right)'&=\yng(1,1,1)\otimes\alpha\ominus(030)\cr
\left(\yng(1,1,1)\otimes\threeplus\right)'
&=\yng(1,1,1)\otimes\threeplus\ominus(040)
}\eqn
$$
It is straightforward to
verify that in the gravitino sequence,
$$
\eqalign{
\left(\;\yng(1,1,1,1)\otimes\alpha\right)'
&=(010)\oplus(101)=\yng(1)\otimes\bar\alpha\;,\cr
\left(\;\yng(1,1,1,1,1)\otimes\alpha\right)'&=(001)=\bar\alpha\;,\cr
\left(\;\yng(1,1,1,1,1,1)\otimes\alpha\right)'&=0\;,\cr
}\eqn
$$
and in the gravity sequence,
$$
\eqalign{
\left(\;\yng(1,1,1,1)\otimes\threeplus\right)'
&=(020)\oplus(100)\oplus(111)\;,\cr
\left(\;\yng(1,1,1,1,1)\otimes\threeplus\right)'&=(011)\;,\cr
\left(\;\yng(1,1,1,1,1,1)\otimes\threeplus\right)'&=0\;,\cr
}\eqn
$$
To be concrete, the exotic graviton has ghosts for ghosts in
\Yboxdim5pt
$\threeplus$, ghosts in $\yng(1)\otimes\threeplus$, gauge potentials
in $\yng(1,1)\otimes\threeplus$ and equations of motion in
$\yng(1,1,1)\otimes\threeplus\ominus(040)$, where
$\yng(2,2,2)_+=(040)$ is the module of the exotic Weyl tensor.
\Yboxdim8pt

The relation to the superfields of
ref. [\ChiodaroliGunaydinRoiban] is not obvious. They
start from representations of the superconformal group. The gauge
invariant fields they list is of course in complete agreement
(scalars in  $(000)(0001)$, spinors in $(010)(0010)$, tensor field
strength in $(020)(0100)$, ``gravitino'' field strength in
$(030)(1000)$ and ``gravity field strength'' in $(040)(0000)$), but it
is not clear how they in turn are built from potentials, since they
appear in a gauge-invariant superfield starting with the scalars. In this
context, the present superspace treatment might be interesting, in
that it displays all the gauge symmetries and fields.
The superfield at ghost number 0 has the scalars at $\theta^2$.


        
\section\ThreeOneSection{$N=(3,1)$ and $N=(3,0)$}One might
expect the construction of superfields for the $N=(3,1)$
model and the non-chiral $N=(2,2)$ supergravity should follow the same
principles, where the field transforms as the ghosts of highest ghost
number.
The pure spinor superfield $\Psi$ for the $N=(3,1)$ model would transform
in the module $(100)(1)$ of the R-symmetry $USp(6)\times USp(2)$,
which is the module of the 12 chiral 2-form fields.
However, I have not been able to perform such a construction with
manifest $N=(3,1)$
supersymmetry, only $N=(3,0)$.

Consider the $N=(3,1)$ supermultiplet. It contains the physical
degrees of freedom\foot{The module $({\bf4},{\bf1},{\bf1},{\bf2})$ is
missing in the list of ref. [\HullExoticGravity].}, labelled by dimensions of
modules of the little group
$SU(2)\times SU(2)\times USp(6)\times USp(2)$,
$$
\matrix{({\bf1},{\bf1},{\bf14}',{\bf2})&&\cr
\downarrow&\searrow&\cr
({\bf2},{\bf1},{\bf14},{\bf2})&&({\bf1},{\bf2},{\bf14}',{\bf1})\cr
\downarrow&\searrow&\downarrow\cr
({\bf3},{\bf1},{\bf6},{\bf2})&&({\bf2},{\bf2},{\bf14},{\bf1})\cr
\downarrow&\searrow&\downarrow\cr
({\bf4},{\bf1},{\bf1},{\bf2})&&({\bf3},{\bf2},{\bf6},{\bf1})\cr
&\searrow&\downarrow\cr
&&({\bf4},{\bf2},{\bf1},{\bf1})\cr
}\eqn
$$
Here, ${\bf14}=(010)$, the $\epsilon$-traceless antisymmetric 2-index
tensor, and ${\bf14'}=(001)$, the $\epsilon$-traceless antisymmetric 3-index
tensor.
Vertical arrows indicate the action of the 3 chiral supersymmetries,
increasing spin, and diagonal arrows of the single anti-chiral
supersymmetry. The two columns form $N=(3,0)$ supermultiplets, which
we may call the (exotic) $N=(3,0)$ gravitino and gravity multiplets.
Note that the full $Spin(1,5)\times USp(6)\times USp(2)$ is unbroken
by this decomposition.

The gravitino multiplet contains the scalars in $({\bf14}',{\bf2})$,
which parametrise the
coset $F_{4(4)}/(USp(6)\times USp(2))$, chiral spinors in
$({\bf14},{\bf2})$, chiral 2-forms in $({\bf6},{\bf2})$ and an chiral exotic
gravitino in $({\bf1},{\bf2})$. The gravity multiplet contains
anti-chiral spinors in $({\bf14}',{\bf1})$, vector fields in
$({\bf14},{\bf1})$, chiral gravitini in $({\bf6}',{\bf1})$
and a ``semi-exotic'' gravity field.

For these two $N=(3,0)$ multiplets it is straightforward to construct
superfields along the same lines as in section \FourZeroSection.
The ``pure spinor'' $\lambda$ transforms in $(010)(100)$
of $Spin(1,5)\times USp(6)$, and obeys
$\lambda^2|_{(100)(000)}=0$. The constraints are again irreducible.

The gravitino multiplet (suppressing the $USp(2)$ doublet, which is
common to all fields in the multiplet) is given by a pure spinor
superfield of ghost number 2 in $(000)(100)=({\bf1},{\bf6})$ 
with a shift symmetry in $(010)(000)$. Its leading component is the
second order ghost for the tensor field.
The gravity multiplet is given by a field of ghost number 1
in $(000)(010)$, with the
ghosts for the gauge fields as leading part.
The zero-mode cohomologies are given in Tables \GravityTable\
and \GravitinoTable.

\Table\GravityTable{$$\hskip-2cm
\vtop{\baselineskip25pt\lineskip0pt
\ialign{
$\hfill#\quad$&$\ss\,\hfill#\hfill\,$&$\ss\,\hfill#\hfill\,$
&$\ss\,\hfill#\hfill\,$&$\ss\,\hfill#\hfill\,$&$\ss\,\hfill#\hfill$
&$\ss\,\hfill#\hfill$&$\ss\,\hfill\qquad#\hfill$\cr
\hfill\hbox{\eightrm gh\#}=&\ts1&\ts0&\ts-1&\ts-2&\ts-3&\ts-4&\cr
&\,\,(000)(010)\,\, \cr
&(010)(100)&\bullet           \cr 
&(020)(000)&(100)(010)&\bullet&       &      \cr
&\bullet&\raise3pt\vtop{\baselineskip6pt\ialign{
					\hfill$#$\hfill\cr
					\ss(001)(001)\cr
					\ss(001)(100)\,(110)(100)\cr}}
                                        &\bullet&\bullet      &\cr
&\bullet&(011)(000)\,(120)(000)&\bullet&\bullet&\bullet\cr
&\bullet&\bullet&\raise3pt\vtop{\baselineskip6pt\ialign{
					\hfill$#$\hfill\cr
					\ss(010)(001)\cr
					\ss(010)(100)\,(101)(100)\cr}}
                 &\bullet&\bullet&\bullet\cr
&\bullet&\bullet&\raise3pt\vtop{\baselineskip6pt\ialign{
					\hfill$#$\hfill\cr
					\ss(020)(000)\,(111)(000)\cr
					\ss(100)(000)\,(100)(010)\cr}}
		&\bullet&\bullet&\bullet\cr
&\bullet&\bullet&\bullet&(001)(100)
                       &\bullet&\bullet\cr
&\bullet&\bullet&\bullet&\raise6pt\vtop{\baselineskip6pt\ialign{
					\hfill$#$\hfill\cr
					\ss(000)(000)\,(000)(010)\cr
					\ss(011)(000)\,(200)(000)\cr}}
                       &\bullet&\bullet\cr
&\bullet&\bullet&\bullet&\bullet&\bullet&\bullet\cr
&\bullet&\bullet&\bullet&\bullet&(100)(000)&\bullet\cr
&\bullet&\bullet&\bullet&\bullet&\bullet&\qquad\bullet\qquad\cr
}}
$$}{The zero-mode cohomology for the $N=(3,0)$
gravity multiplet.}

\Table\GravitinoTable{$$\hskip-2cm
\vtop{\baselineskip25pt\lineskip0pt
\ialign{
$\hfill#\quad$&$\ss\,\hfill#\hfill\,$&$\ss\,\hfill#\hfill\,$
&$\ss\,\hfill#\hfill\,$&$\ss\,\hfill#\hfill\,$&$\ss\,\hfill#\hfill$
&$\ss\,\hfill#\hfill$&$\ss\,\hfill#\hfill$&\quad#\cr
\hfill\hbox{\eightrm gh\#}=&\ts2&\ts1&\ts0&\ts-1&\ts-2&\ts-3&\ts-4&\cr
&\,\,(000)(100)\,\, \cr
&(010)(000)&\bullet&             \cr 
&\bullet&(100)(100)&\bullet&       &      &\cr
&\bullet&(001)(000)\,(110)(000)&\bullet&\bullet&       &\cr
&\bullet&\bullet&(000)(001)\,(011)(100)&\bullet\cr
&\bullet&\bullet&\raise3pt\vtop{\baselineskip6pt\ialign{
					\hfill$#$\hfill\cr
					\ss(010)(000)\,(010)(010)\cr
					\ss(021)(000)\,(101)(000)\cr}}
                 &\bullet&\bullet&\bullet\cr
&\bullet&\bullet&\bullet
		&(002)(100)&\bullet&\bullet&\bullet\cr
&\bullet&\bullet&\bullet&\raise3pt\vtop{\baselineskip6pt\ialign{
					\hfill$#$\hfill\cr
					\ss(001)(000)\,(001)(010)\cr
					\ss(012)(000)\,(110)(000)\cr}}
                       &\bullet&\bullet&\bullet\cr
&\bullet&\bullet&\bullet&(000)(001)
                       &\bullet&\bullet&\bullet\cr
&\bullet&\bullet&\bullet&\bullet&(010)(000)\,(101)(000)&\bullet&\bullet\cr
&\bullet&\bullet&\bullet&\bullet&\bullet&\bullet&\bullet\cr
&\bullet&\bullet&\bullet&\bullet&\bullet&(001)(000)&\bullet\cr
&\bullet&\bullet&\bullet&\bullet&\bullet&\bullet&\qquad\bullet\qquad\cr
}}
$$}{The zero-mode cohomology for the $N=(3,0)$
gravitino multiplet.}

From the tables, we read the sequences of ghosts fields and verify that the
description is correct. The new information is the sequence 
for the ``semi-exotic'' $N=(3,1)$ graviton, which becomes
$$
\threeplus
\rightarrow\yng(1)\otimes\threeplus
\rightarrow\yng(1,1)\otimes\threeplus\ominus(031)
\rightarrow\left(\;\yng(1,1,1)\otimes\threeplus\right)'
\rightarrow\left(\;\yng(1,1,1,1)\otimes\threeplus\right)'
\eqn
$$
where $()'$ now indicates the removal of modules coming through
\Yboxdim5pt
$\yng(2,2,1)_+=(031)$, the module of the ``semi-exotic Weyl tensor''.
\Yboxdim8pt

Going back to $N=(3,1)$, one might have suspected that the
cohomologies of the $N=(3,0)$ gravity and gravitino multiplets would
combine into the cohomology of an $N=(3,1)$ superfield transforming in
$(000)(100)(1)$. One should then use a ``pure spinor'' $\lambda$
in $(010)(100)(0)\oplus(001)(000)(1)$.
Surprisingly, it turns out that such a field seems to become ``topological'',
in the sense that all sequences of zero-mode cohomologies (ghosts and
component fields) consist of the tensor product of some modules (the
ghost number 2 ghosts of the $N=(3,0)$ gravitino multiplet) with forms of
increasing degree, and no local degrees of freedom remain.
I have no clear explanation of this. An investigation of $N=(2,2)$
supergravity might provide some insight.

\section\Conclusions{Conclusions}This note has presented superfields
for the $N=(4,0)$ and $(3,1)$ exotic supergravity multiplets in
$D=6$, in the latter case in terms of $N=(3,0)$ superfields.

It is not be possible to write a covariant action reproducing
$Q\Psi=0$ (only). Non-covariant component actions are given in ref.
[\HenneauxLekeuLeonard,\HenneauxLekeuMatulichProhaska].
A pseudo-action, where some fields need to be
(consistently) set to 0, should exist. Then one would first need
to extend the pure spinor superspace to include non-minimal variables
[\BerkovitsNonMinimal]. This enables non-degenerate integration, and
the procedure is standard [\PureSpinorOverview]; the integration
measure corresponds to the top cohomology of the singlet superfield
of eq. (\ZeroModeForms). One would also need a
``conjugate'' pure spinor superfield $\Psi^\star$, which
(for the $N=(4,0)$ model) would
presumably transform in the module $(011)(0000)$, since this is the
highest cohomology in $\Psi$. The
cohomology in $\Psi^\star$ is conjugate to the one in $\Psi$. And, it
is necessary to show that integration, with some $\lambda$'s inserted,
implies the proper shift symmetries of $\Psi$ and $\Psi^\star$. All this
remains yet to be done, but it is doubtful that it would
contribute significantly to the
understanding of the model. The corresponding
procedure for the $N=(2,0)$ tensor multiplet works
[\CederwallTensorAction], and yields two tensor multiplets of the
same chirality, of which one can be set to zero.

Also without a pseudo-action, one may discuss possible deformations of
$Q\Psi=0$. Such  considerations are usually
especially powerful in a pure spinor
superfield framework, since all gauge symmetries and equations are
treated simultaneously, and the field carries as low dimension and as
high ghost number as possible. However, na\"\i ve dimensional
considerations point at the impossibility of finding local
interactions this way, just as they do for the $N=(2,0)$ theory
(without the introduction of some dimensionful constant).

\vfill\eject

\acknowledgements The author would like to thank
Marc Henneaux, Victor Lekeu, Jakob
Palmkvist and Ingmar Saberi for discussions on various aspects of the
models and the formalism.


\refout

\end